\documentclass[useAMS, usenatbib, amssymb]{mn2e}
\usepackage{psfig}
\usepackage{graphicx}
\usepackage{epsf}
\usepackage{bm}

\title [LSR from RAVE DR3]
{Local Stellar Kinematics from RAVE Data\\
I. Local Standard of Rest}

\author[Co\c skuno\u glu et al.]
       {B. Co\c skuno\u glu$^{1}$\thanks{E-mail: basarc@istanbul.edu.tr},
        S. Ak$^{1}$, S. Bilir${^1}$, S. Karaali$^{2}$, E. Yaz$^{1}$, G. Gilmore$^{3}$, G. M. Seabroke$^{4, 5}$,
\newauthor
O. Bienaym\' e$^{6}$, J. Bland-Hawthorn$^{7}$, R. Campbell$^{8}$, K. C. Freeman$^{9}$, B. Gibson$^{10}$,  
\newauthor
E. K. Grebel$^{11}$, U. Munari$^{12}$, J. F. Navarro$^{13}$, Q. A. Parker$^{14, 16}$, A. Siebert$^{6}$,
\newauthor
A. Siviero$^{12, 15}$, M. Steinmetz$^{15}$, F. G. Watson$^{16}$, R. F. G. Wyse$^{17}$, T. Zwitter$^{18, 19}$
\\
  $^1$Istanbul University, Science Faculty, Department of Astronomy and Space 
Sciences, 34119, University-Istanbul, Turkey\\
  $^2$Beykent University, Faculty of Science and Letters, Department of Mathematics  
and Computer, Beykent, 34398, Istanbul, Turkey\\
  $^3$Institute of Astronomy, Madingley Road, Cambridge, CB3 OHA, UK\\
  $^4$e2v Centre for Electronic Imaging, Planetary and Space Sciences Research Institute, The Open University, Milton Keynes, UK\\
  $^5$Mullard Space Science Laboratory, University College London, Hombury St Mary, Dorking, RH5 6NT, UK\\
  $^6$Observatoire de Strasbourg, Strasbourg, France\\
  $^7$Sydney Institute for Astronomy, School of Physics, University of Sydney, NSW 2006, Australia\\
  $^8$Western Kentucky University, Bowling Green, Kentucky, USA\\
  $^9$RSAA, Australian National University, Canberra, Australia\\
  $^{10}$University of Central Lancashire, Preston, UK\\
  $^{11}$Astronomisches Rechen-Institut, Zentrum f\" ur Astronomie der Universit\"at Heidelberg, Heidelberg, Germany\\
  $^{12}$INAF Astronomical Observatory of Padova, 36012 Asiago (VI), Italy\\
  $^{13}$Department of Physics and Astronomy, University of Victoria, Victoria, BC, Canada V8P 5C2\\
  $^{14}$Macquarie University, Sydney, Australia\\
  $^{15}$Astrophysikalisches Institut Potsdam, Potsdam, Germany\\ 
  $^{16}$Australian Astronomical Observatory, Sydney, Australia\\
  $^{17}$John Hopkins University, Baltimore, Maryland, USA\\
  $^{18}$University of Ljubljana, Faculty of Mathematics and Physics, Ljubljana, Slovenia\\
  $^{19}$Center of Excellence SPACE-SI, Ljubljana, Slovenia.\\} 
  
\date{}

\pagerange{\pageref{firstpage}--\pageref{lastpage}} \pubyear{}

\begin{document}

\maketitle

\label{firstpage}

\begin{abstract}
We analyze a sample of 82850 stars from the RAVE survey, with 
well-determined velocities and stellar parameters, to isolate a 
sample of 18026 high-probability thin-disc dwarfs within 600 pc of 
the Sun. We derive space motions for these stars, and deduce the 
solar space velocity with respect to the Local Standard of Rest.
The peculiar solar motion we derive is in excellent agreement in
radial $U_{\odot}$ and vertical $W_{\odot}$ peculiar motions with other
recent determinations. Our derived tangential peculiar
velocity, $V_{\odot}$ agrees with very recent determinations, which
favour values near 13 km s$^{-1}$, in disagreement with earlier
studies. The derived values are not significantly dependent on the
comparison sample chosen, or on the method of analysis. The local
galaxy seems very well dynamically relaxed, in a near symmetric potential.

\end{abstract}


\section{Introduction}

The kinematics of stars near the Sun
provide vital information regarding the structure and the evolution of
the Galaxy. Since the observations are relative to the Sun, which is
not stationary, the kinematical parameters we obtain need to be
corrected accordingly to the Local Standard of Rest (LSR), which is
defined as the rest frame of a star at the location of the Sun that
would be on a circular orbit in the gravitational potential of the
Galaxy. This implies the determination of the solar peculiar velocity
components relative to that standard, $U_{\odot}$, $V_{\odot}$ and
$W_{\odot}$. These values are also of considerable intrinsic interest, as
their determination as a function of the spatial distribution and
kinematic properties of the comparison stellar sample is a test of the
symmetry of the local Galactic potential as a function of, for
example, velocity dispersion. One can test if thin disk slowly moving
stars are, or are not, affected by a Galactic bar, among many other 
considerations. Uncertainty in the solar kinematic
parameters can have profound implications for Galactic structural
analyses \citep[cf. e.g.][]{McMB2010} and may be affected systematically
in turn by presumptions about larger scale Galactic stellar
populations \citep[cf. e.g.][]{Schonrich10}. There have been several very
recent determinations of the solar peculiar velocity, including some
using RAVE (RAdial Velocity Experiment, Steinmetz et al., 2006, see also 
Section 2) data as we use here \citep[e.g.][]{Pasetto10}. A 
summarized list of recent determinations of the space velocity components
of the Sun appearing in the literature is given in Table 1, while
\cite{Francis09} provide an even longer list of references. Given the
astrophysical significance of systematic sample- and method-dependent
effects in the determination, we consider here a further determination
of the solar space velocity, using a complementary technique.

The determination of the solar space motion relative to a Local Standard
of Rest (LSR) is a long-standing challenge. The robust determination of 
the systematic motion of some stellar population relative to some agreed
LSR would itself be of considerable interest. Hence many studies have
been carried out since that of \cite{Homann86} to estimate these
parameters. However, there is still uncertainty about the numerical
values of the solar velocity components, relative to local thin disc
stars, especially for the $V_{\odot}$ component, due to the complexity
involved in compensating for the velocity ``lag'' of sample stars. The
difference between the results of different researchers originates
from different data as well as different procedures used. The very
different results derived by \cite{Dehnen98} and \cite{Schonrich10},
both of whom who used the {\em Hipparcos} data, are an illustration of
the continuing model-dependence of the result.

The radial ($U_{\odot}$) and vertical ($W_{\odot}$) components of the
solar motion, as a function of the relevant stellar population, can be
obtained in principle by direct analysis of the mean heliocentric
velocities of solar neighbourhood stars, subject to appropriate sample
selection. As one can see in Table~1, the various determinations are
in tolerable agreement. However, the determination of the component in
the direction of Galactic rotation ($V_{\odot}$) is complicated by
systematics: the mean lag, i.e. asymmetric drift $V_{a}$, with respect
to the LSR which depends on the velocity dispersion ($\sigma$) of the
stellar sample in question. An extensive discussion of the radial
Jeans' equation on which analysis of the asymmetric drift is based,
including its derivation, relevant approximations, and applications,
can be found in \cite{GWK89}. At its simplest, a single population
asymmetric drift/Jeans' analysis assumes a linear relation between
(suitable values of) the asymmetric drift $V_{a}$ of any stellar
sample and its squared radial velocity dispersion $\sigma^2_R$
\citep*{Stromberg46}. Formal least-squares straight line fits
intercept the $V_{a}$ axis at some negative value, typically in the
range -20 to -5 km s$^{-1}$. This $V_{a}$ intercept determines the
solar velocity $V_{\odot}$ using the equation
$\bar{V_s}$=$V_{a}$+$V_{\odot}$. $\bar{V_s}$ is the negative mean
heliocentric azimuthal velocity of any stellar sample. The value
$V_{\odot}$=5.25$\pm$0.62 km s$^{-1}$ of \cite{Dehnen98} (hereafter
DB98) is typical of the smaller values deduced recently from this
procedure.

However, four other recent studies argue for larger values of
$V_{\odot}$ than that of DB98. \cite{Piskunov06} determined a value
different from that of DB98 by about 7 km s$^{-1}$ in the $V$
component of the solar motion, when determined with respect to open
clusters in the solar neighbourhood. \cite{Binney10} fitted
distribution function models to two sets of velocity distributions and
obtained a difference of about 6 km s$^{-1}$ compared to the value of
DB98. This higher value of $V_{\odot}$ has been confirmed by
\cite{Reid09} and \cite{Rygl10} in their works related to radio
frequency astrometry of masers in regions of massive star formation
(though see \cite{McMB2010} for a different analysis).
\cite{Schonrich10} applied a particular chemodynamical model of the
Galaxy to DB98's data and determined a value for the $V_{\odot}$ solar
velocity component consistent with recent high values. The key factor
in the \cite{Schonrich10} analysis, and of others, is the relevant
range of radial velocity dispersions $\sigma^2_R$ which is used in the
Jeans' analysis. \cite{Schonrich10} showed that for
${\sigma}^2_R$$\geq$600 (km s$^{-1}$)$^2$ the {\em Hipparcos} data
define a straight line, whereas for ${\sigma}^2_R$$\leq$400 (km
s$^{-1}$)$^2$ they deviate from such a simple fit. If one omits stars
with low dispersions, and extrapolates the linear fit for stars with 
high dispersion to zero, the linear fit for
stars with high dispersions, one confirms the low value of
DB98. Actually, this is the procedure used by DB98, who also ignored
stars with low velocity dispersion due to their probable lack of
dynamical equilibrium. While it is true that low velocity dispersion 
stars are likely to be most affected by the effects of dissolving 
star clusters and the non-axisymmetric gravitational potential 
of spiral arms, one does need to consider carefully the offset 
character of the relation between $V_{a}$ and ${\sigma}^2_R$ 
at low velocity dispersions to avoid bias.

Here we use a new sample of stars and a different methodology to
re-estimate the $U_{\odot}$, $V_{\odot}$ and $W_{\odot}$ solar
velocity components. 1) We use RAVE data extending to larger
distances than the {\em Hipparcos} sample. 2) We applied the
following constraints to obtain a sample of main sequence stars: i) we
selected stars with surface gravity 4$<$$\log g$$<$5, ii) we omitted
stars with $(J-H)_0$$<$0.05 and $(J-H)_0$$>$0.55 to avoid the blue
horizontal branch and possible red giant stars, iii) we excluded stars
with space velocity errors larger than 25 km s$^{-1}$, iv) we
separated stars into populations (see Section 5) and we used the thin 
disc population which is not contaminated by thick disc/halo as a 
preferred sample in our work. The last constraint is especially 
important in estimating $V_{\odot}$, because it excludes thick disc 
and halo stars which have relatively large asymmetric drift. Thus, 
it limits the range of $V$ velocity component, which in turn 
minimizes possible astrophysical large-scale dynamical asymmetry effects.

The RAVE survey is described in Section 2, the data are presented in
Section 3, Sections 4 and 5 are devoted to kinematics and population
analysis, respectively. The results are given in Section 6 and
finally a discussion is presented in Section 7.

\begin{table*}
\center 
\caption{Space velocity components of the Sun with respect to the LSR as given in the literature.}
\begin{tabular}{llccc}
\hline
Reference & Source &       $U_{\odot}$ &   $V_{\odot}$    &  $W_{\odot}$ \\
       &  &     (km s$^{-1}$)&   (km s$^{-1}$) & (km s$^{-1}$) \\
\hline
This study (2010) & RAVE DR3 &8.50$\pm$0.29 &  13.38$\pm$0.43 &  6.49$\pm$0.26 \\
Bobylev \& Bajkova (2010)& Masers & 5.5$\pm$2.2 & 11.0$\pm$1.7 & 8.5$\pm$1.2 \\
Breddels et al. (2010)& RAVE DR2 & 12.0$\pm$0.6 & 20.4$\pm$0.5 & 7.8$\pm$0.3 \\
Sch\"onrich et al. (2010)& Hipparcos & 11.10$\pm$0.72 &  12.24$\pm$0.47 &  7.25$\pm$0.36 \\
Francis \& Anderson (2009)& Hipparcos & 7.5$\pm$1.0 &  13.5$\pm$0.3 &  6.8$\pm$0.1 \\
Veltz et al. (2008)& RAVE DR1 & 8.5$\pm$0.3 & -- &  11.1$\pm$1.0 \\
Bobylev \& Bajkova (2007)& F \& G dwarfs & 8.7$\pm$0.5 &  6.2$\pm$2.2 &  7.2$\pm$0.8 \\
Piskunov et al. (2006)& Open clusters& 9.44$\pm$1.14 & 11.90$\pm$0.72 & 7.20$\pm$0.42 \\
Mignard (2000)& K0-K5 &       9.88 &      14.19 &       7.76 \\
Dehnen \& Binney (1998)& Hipparcos, d$_{max}$=100 pc & 10.00 $\pm$ 0.36 & 5.25 $\pm$ 0.62 & 7.17 $\pm$ 0.38 \\
Binney et al. (1997)& Stars near South Celestial Pole &   11 $\pm$ 0.6 &  5.3 $\pm$ 1.7 &  7.0 $\pm$ 0.6 \\
Mihalas \& Binney (1981)& Galactic Astronomy 2$^{nd}$ Ed. &  9.2 $\pm$ 0.3 &         12.0 &  6.9 $\pm$ 0.2 \\
Homann (1886)& Solar neighbourhood stars & 17.4$\pm$11.2 & 16.9$\pm$10.9 & 3.6$\pm$2.3 \\
\hline
\end{tabular} 
\end{table*}

\section{RAVE}

The RAdial Velocity Experiment \citep[RAVE,][]{Steinmetz06} is a
spectroscopic survey aiming to measure radial velocities and stellar
atmospheric parameters, temperature, metallicity, surface gravity, of
up to one million stars using the 6 degree Field (6dF) multi-object
spectrograph on the 1.2 m UK Schmidt Telescope of the Australian Astronomical 
Observatory. The RAVE programme started in 2003, obtaining medium
resolution spectra in the Ca-triplet region (8410-8795 \AA) for
southern hemisphere stars in the magnitude range
9$<$$I_{DENIS}$$<$13. The scientific goals of RAVE include analyzing
the chemical and dynamical evolution of the Galaxy, using both dwarfs
and giants observed locally. The main sequence stars occupy a region
extending to a distance of a few hundred parsecs, whereas giants
extend up to a few kpc. RAVE was designed to avoid high reddening in
the Galactic plane and contamination from the bulge,
i.e. $\mid$$b$$\mid$$>$5$^{\circ}$ and $l$$<$315$^{\circ}$ constraints
are applied when selecting programme stars.
 
RAVE is a precursor
of the spectroscopic part of the cornerstone mission Gaia of the
European Space Agency. The wavelength range for RAVE spectra was
chosen to match that of the Gaia Radial Velocity Spectrometer
\citep{Munari03, Katz04, Wilkinson05}, i.e. around the Ca II IR
triplet. This wavelength range also includes lines from elements 
such as Fe, Ca, Si, Mg and Ti which can be used to estimate 
[$\alpha$/Fe] in addition to overall metallicity 
\citep[see][for a more detailed description of the goals of RAVE]
{Steinmetz06}.

\section{Data}

The data used in this study are a working version of what will become
RAVE's third data release (DR3; Siebert et al., in prep.). DR3 will
consist of 82850 stars, each with equatorial and Galactic
coordinates, radial velocity, metallicity, surface temperature and 
surface gravity. We also note the two existing
data releases, i.e. DR1 \citep{Steinmetz06},
DR2 \citep{Zwitter08}. Proper motions were compiled from several
catalogues: {\em Tycho-2}, Supercosmos Sky Survey, PPMXL and
UCAC-2. Proper motion accuracy decreases in this order, therefore, 
if proper motions were available from all catalogues, {\em Tycho-2}'s 
value was used. If {\em Tycho-2} did not provide proper motions the 
values were taken from Supercosmos Sky Survey, etc. Photometric data 
are based on optical and near infrared (NIR) systems. The magnitudes of 
stars were obtained by matching RAVE DR3 with {\em Tycho-2}, USNO-B, 
DENIS and {\em 2MASS} catalogues. The analysis here uses stars in RAVE 
DR3 with {\em 2MASS} catalogue photometry.

\subsection{Main Sequence Stars and Distance Determination}

We applied two constraints to obtain a main sequence sample: we
selected stars with surface gravities 4$<$$\log g$$<$5 and we excluded
stars with $(J-H)_0$$<$0.05 and $(J-H)_0$$>$0.55. The second
constraint is especially effective in reducing the contamination due
to blue horizontal branch and red giant stars. Thus, the sample was
reduced to 21310 stars. The $(J-H)_0$-$(H-K_s)_0$ two colour diagram
of the parent sample (Fig.~1a) has a bimodal distribution, confirming
the presence of K and M giants in addition to the F-M dwarfs, whereas
the sample obtained after applying the two constraints cited, shown in
Fig.~1b, has only one mode indicating a pure main sequence sample.

\begin{figure}
\begin{center}
\includegraphics[scale=0.35, angle=0]{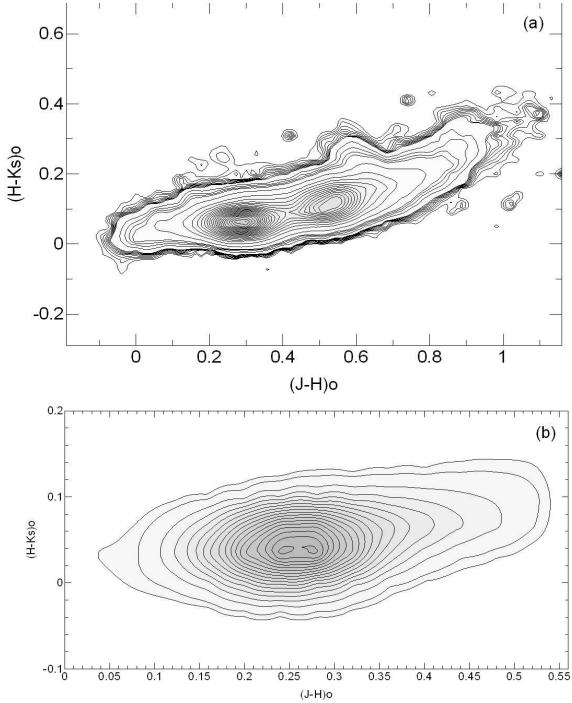}
\caption[] {Two colour diagrams for two samples of RAVE DR3 stars. (a)
  for the original sample (main sequence stars and giants), (b) for
  main sequence stars only.}
\end{center}
\end{figure}

\begin{figure*}
\begin{center}
\includegraphics[scale=0.75, angle=0]{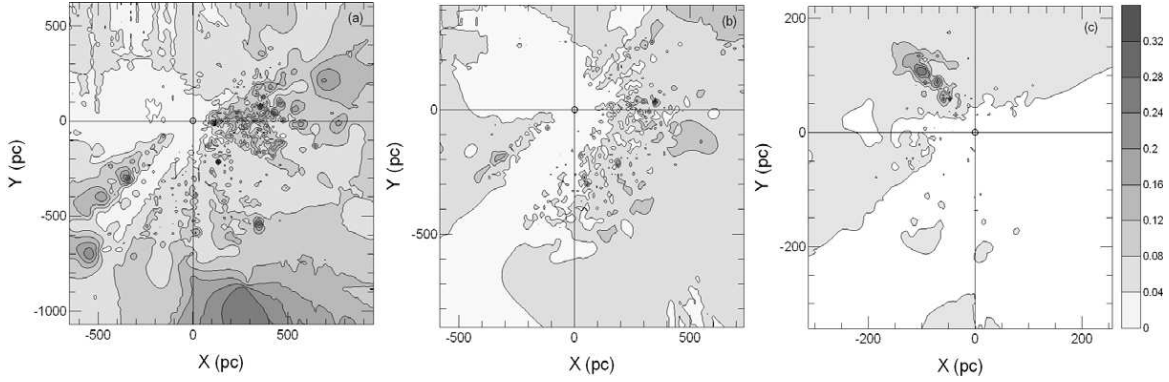}
\caption[] {$E(B-V)$ reddening map in the solar neighbourhood for our RAVE main 
sequence sample. (a) stars with $|b|\leq30^\circ$, (b) stars with 
$30^\circ<|b|\leq60^\circ$ and (c) stars with $|b|>60^\circ$.}
\end{center}
\end{figure*}

Contrary to the {\em Hipparcos} catalogue, parallaxes are not
available for stars observed in the RAVE survey, but not existing in the Hipparcos 
survey. Hence, the distances of the main sequence stars were calculated using 
another procedure. We applied the main-sequence colour-luminosity relation of
\cite{Bilir08}, which, as described in that reference, is valid in the
absolute magnitude range 0$<$$M_J$$<$6. The errors of the distances were estimated 
combining the internal errors of the coefficients of Bilir et al.'s (2008) equation and 
the errors of the {\em 2MASS} colour indices. 

As most of the stars in the
sample are at distances larger than 100 pc, their colours and
magnitudes will be affected by interstellar reddening. Hence,
distance determination is carried out simultaneously with de-reddening
of the sample stars.  As the first step in an iterative process, we
assume the original $(J-H)$ and $(H-K_s)$ colour indices are
de-reddened, and evaluate the $M_J$ absolute magnitudes of the sample
stars by means of the colour-luminosity relation of
\cite{Bilir08}. Combination of the apparent and absolute magnitudes
for the $J$ band gives the distance $d$ of a star. We used the maps of
\cite*{Schlegel98} and evaluated the colour excess $E_{\infty}(B-V)$ for each
sample star. The relation between the total and selective absorptions
in the $UBV$ system, i.e.
 
\begin{equation}
A_{\infty}(b)=3.1E_{\infty}(B-V)
\end{equation}
gives $A_{\infty}(b)$ which can be used in evaluating $A_d(b)$ 
using Bahcall \& Soneira's (1980) procedure:

\begin{equation}
A_{d}(b)=A_{\infty}(b)\Biggl[1-\exp\Biggl(\frac{-\mid
d~\sin(b)\mid}{H}\Biggr)\Biggr]
\end{equation}
where $b$ and $d$ are the Galactic latitude and distance of the star,
respectively. $H$ is the scaleheight for the interstellar dust which
is adopted as 125 pc \citep{Marshall06} and $A_{\infty}(b)$ and
$A_{d}(b)$ are the total absorptions for the model and for the
distance to the star, respectively. Then, the colour excess at the
distance of the star, $E_{d}(B-V)$, can be evaluated using a specific form of
Eq. 1:

\begin{equation}
A_{d}(b)=3.1E_{d}(B-V)
\end{equation}
That value was used in Fiorucci \& Munari's (2003) equations to obtain
the total absorptions for the $J$, $H$ and $K_s$ bands, i.e. $A_J$ =
0.887 $\times$ $E(B-V)$, $A_H$ = 0.565 $\times$ $E(B-V)$ and $A_{Ks}$
= 0.382 $\times$ $E(B-V)$, which were used in Pogson''s equation 
($m_{i}-M_{i}=5\log d-5+A_{i}$; $i$ denotes a specific band)  to
evaluate distances. Contrary to the assumption above, the original
$(J-H)$ and $(H-K_s)$ colour indices are not de-reddened. Hence, the
application of the equations (1) to (3) is iterated until the
distance $d$ and colour index $E_{d}(B-V)$ approach constant values.

Our resulting distribution of $E(B-V)$ reddening for our 21310 RAVE main
sequence stars in the $X$–-$Y$ Galactic plane is given in
Fig. 2. To analyze the reddening in the solar neighbourhood more accurately, 
we divided the stars into three subsamples according to their Galactic latitude: 
Fig. 2a shows the stars with $|b|\leq30^\circ$, in Fig. 2b the stars with 
$30^\circ<|b|\leq60^\circ$ are shown, whereas Fig. 2c gives the stars with 
$|b|>60^\circ$. The first feature of the reddening distribution is the complex
structure of the reddening in the first two panels. The local bubble, i.e. the region 
within 0.1 kpc distance from the Sun, is not affected by the reddening effect
of the interstellar dust, whereas $E(B-V)$ can be as high as 0.35 mag
at larger distances. As expected high latitude stars (Fig. 2c) have smaller 
reddening values. The second feature is that one can not ignore the
interstellar reddening even when using NIR bands.  

The distribution of distances (Fig. 3) shows that 80\% of the sample
stars have almost a normal distribution within the distance interval
0$\leq$$d$$\leq$0.4 kpc, whereas the overall distribution which
extends up to 1 kpc is skewed, with a median of 0.276 kpc. However,
97\% of the sample stars are within $d$=0.6 kpc.

\begin{figure}
\begin{center}
\includegraphics[scale=0.35, angle=0]{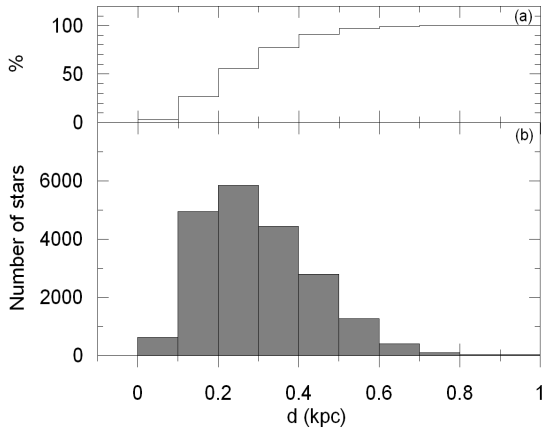}
\caption[] {Cumulative (a) and frequency (b) distributions of
  distances of main sequence stars.}
\end{center}
\end{figure}

The position of the sample stars in the rectangular coordinate system
relative to the Sun is given in Fig. 4. The projected shapes both on
the Galactic ($X$, $Y$) plane, and on the vertical ($X$, $Z$) plane of
the sample show asymmetrical distributions. The median coordinates
($X$=60, $Y$=-107, $Z$=-108 pc) of the sample stars confirm this
appearance. The inhomogeneous structure is due to the incomplete
observations of the RAVE project and that the programme stars were
selected from the Southern Galactic Hemisphere.  

\begin{figure}
\begin{center}
\includegraphics[scale=0.35, angle=0]{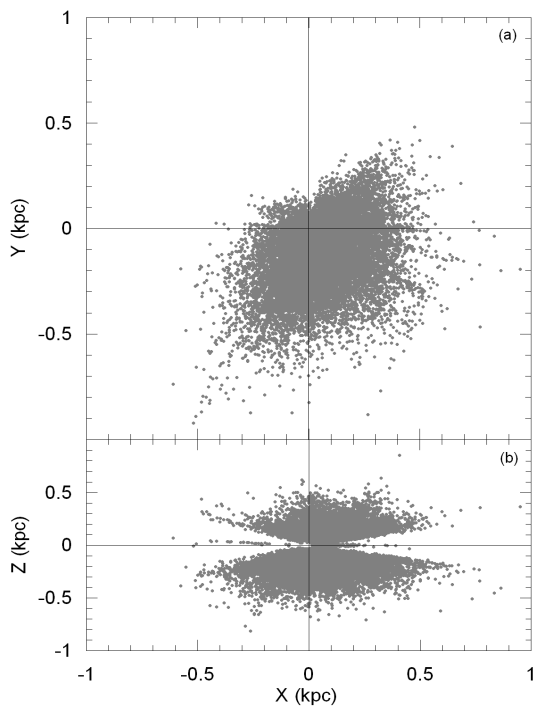}
\caption[] {Space distributions of RAVE main sequence stars on two
  planes. (a) $X$-$Y$ and (b) $X$-$Z$.}
\end{center}
\end{figure}

\section{Kinematics}

We combined the distances estimated in Section 3 with RAVE kinematics
and the available proper motions, 
applying the (standard) algorithms and the transformation matrices of
\cite{Johnson87} to obtain their Galactic space velocity components
($U$, $V$, $W$). In the calculations, the epoch of J2000 was adopted
as described in the International Celestial Reference System (ICRS) of
the {\em Hipparcos} and {\em Tycho-2} Catalogues \citep{ESA97}. The
transformation matrices use the notation of a right handed
system. Hence, $U$, $V$ and $W$ are the components of a velocity
vector of a star with respect to the Sun, where $U$ is positive
towards the Galactic centre ($l$=0$^{\circ}$, $b$=0$^{\circ}$), $V$ is
positive in the direction of Galactic rotation ($l$=90$^{\circ}$,
$b$=0$^{\circ}$) and $W$ is positive towards the North Galactic Pole
($b$=90$^{\circ}$).

Correction for differential Galactic rotation is necessary for
accurate determination of $U$, $V$ and $W$ velocity components. The
effect is proportional to the projection of the distance to the stars
onto the Galactic plane, i.e. the $W$ velocity component is not
affected by Galactic differential rotation \citep{Mihalas81}. We
applied the procedure of \cite{Mihalas81} to the distribution of the
sample stars in the $X$–-$Y$ plane and estimated the first order
Galactic differential rotation corrections for $U$ and $V$ velocity
components of the sample stars. The range of these corrections is
-25.13$<$$dU$$<$13.07 and -1.46$<$$dV$$<$2.28 km s$^{-1}$ for $U$ and
$V$, respectively. As expected, $U$ is affected more than the $V$
component. Also, the high values for the $U$ component show that
corrections for differential Galactic rotation can not be ignored.

The uncertainty of the space velocity components $U_{err}$, $V_{err}$
and $W_{err}$ were computed by propagating the uncertainties of the
proper motions, distances and radial velocities, again using a
(standard) algorithm by \cite{Johnson87}. Then, the error for the
total space motion of a star follows from the equation:

\begin{equation}
S_{err}^{2}=U_{err}^{2}+V_{err}^{2}+W_{err}^{2}. 
\end{equation}

The distributions of errors for both the total space motion and that
in each component are plotted in Fig. 5. The median and standard
deviation for space velocity errors are \~ S$_{err}$=8.01 km s$^{-1}$
and $s$=16.99 km s$^{-1}$, respectively.  We now remove the most
discrepant data from the analysis, knowing that outliers in a survey
such as this will preferentially include stars which are systematically
mis-analysed binaries, etc. Thus, we omit stars with errors that
deviate by more than the sum of the standard error and the standard 
deviation, i.e. $S_{err}$$>$25 km s$^{-1}$. This removes 867 stars, 
4\% of the sample. Thus, our sample was reduced to
20453 stars, those with more robust space velocity components. After
applying this constraint, the median values and the standard
deviations for the velocity components were reduced to (\~U$_{err}$,
\~V$_{err}$, \~W$_{err}$)=(4.83$\pm$3.19, 4.40$\pm$2.82,
3.94$\pm$2.69) km s$^{-1}$. The two dimensional distribution of the
velocity components for the reduced sample is given in Fig. 6. The
most interesting feature for our present analysis in the ($U$, $V$) and
($V$, $W$) diagrams is the offsets of the zero points of the sample
stars from the origins of the coordinate systems - these correspond to
the solar velocity components.

\begin{figure}
\begin{center}
\includegraphics[scale=0.45, angle=0]{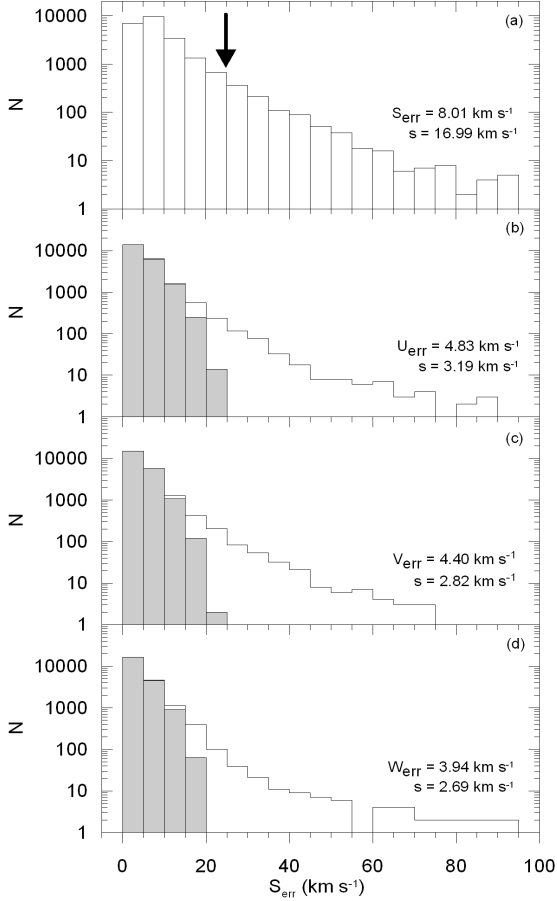}
\caption[] {Error histograms for space velocity (a) and its components
  (b-d) for RAVE main sequence stars. The arrow in panel (a)
  indicates the upper limit of the total error adopted in this work. 
  The shaded part of the histogram indicates the error for different 
   velocity components of stars after removing the stars with large 
   space velocity errors.}
\end{center}
\end{figure}

\begin{figure}
\begin{center}
\includegraphics[scale=0.45, angle=0]{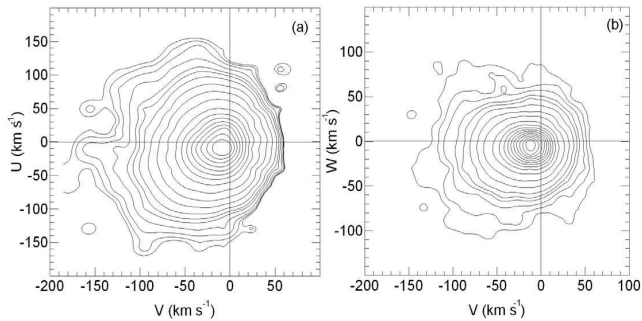}
\caption[] {The distribution of velocity components of RAVE main
  sequence stars in two projections onto the Galactic Plane: (a)
  $U$-$V$ and (b) $W$-$V$.}
\end{center}
\end{figure}

\section{Population Analysis} 

We now wish to consider the population kinematics as a function of
stellar population, using space motion as a statistical process to
label stars as (probabilistic) ``members'' of a stellar population. We
used the procedure of \cite*{Bensby03} and \cite{Bensby05} to allocate
the main sequence sample (20453 stars) into populations and derived
the solar space velocity components for the thin disc population to
check the dependence of LSR parameters on population. \cite{Bensby03,
Bensby05} assumed that the Galactic space velocities of stellar
populations with respect to the LSR have Gaussian distributions as
follows:

\begin{equation}
f(U,~V,~W)=k~.~\exp\Biggl(-\frac{U_{LSR}^{2}}{2\sigma_{U{_{LSR}}}^{2}}-\frac{(V_{LSR}-V_{asym})
^{2}}{2\sigma_{V{_{LSR}}}^{2}}-\frac{W_{LSR}^{2}}{2\sigma_{W{_{LSR}}}^{2}}\Biggr),
\end{equation}
where 
\begin{equation}
k=\frac{1}{(2\pi)^{3/2}\sigma_{U{_{LSR}}}\sigma_{V{_{LSR}}}\sigma_{W{_{LSR}}}}
\end{equation}
normalizes the expression. For consistency with other analyses, we
adopt $\sigma_{U{_{LSR}}}$, $\sigma_{V{_{LSR}}}$ and
$\sigma_{W{_{LSR}}}$ as the characteristic velocity dispersions: 35,
20 and 16 km s$^{-1}$ for thin disc ($D$); 67, 38 and 35 km s$^{-1}$
for thick disc ($TD$); 160, 90 and 90 km s$^{-1}$ for halo ($H$),
respectively \citep{Bensby03}. $V_{asym}$ is the asymmetric drift:
-15, -46 and -220 km s$^{-1}$ for thin disc, thick disc and halo,
respectively. $U_{LSR}$, $V_{LSR}$ and $W_{LSR}$ are LSR
velocities. The space velocity components of the sample stars relative
to the LSR were estimated by adding the values for the space velocity
components evaluated by \cite{Dehnen98} to the corresponding solar ones.

The probability of a star of being ``a member'' of a given population is
defined as the ratio of the $f$($U$, $V$, $W$) distribution functions
times the ratio of the local space densities for two
populations. Thus,
 
\begin{equation}
TD/D=\frac{X_{TD}}{X_{D}}.\frac{f_{TD}}{f_{D}}~~~~~~~~~~TD/H=\frac{X_{TD}}{X_{H}}.\frac{f_{TD}}{f_{H}}
\end{equation}
are the probabilities for a star of it being classified as a thick
disc star relative to being a thin disc star, and relative to
being a halo star, respectively. $X_{D}$, $X_{TD}$ and $X_{H}$ are
the local space densities for thin disc, thick disc and halo,
i.e. 0.94, 0.06, and 0.0015, respectively \citep{Robin96,Buser99}. We
followed the argument of \cite{Bensby05} and separated the sample
stars into four categories: $TD/D$$\leq$0.1 (high probability thin
disc stars), 0.1$<$$TD/D$$\leq$1 (low probability thin disc stars),
1$<$$TD/D$$\leq$10 (low probability thick disc stars) and $TD/D$$>$10
(high probability thick disc stars). It turned out that 18026 and 1552
stars of the sample were classified as high and low probability thin
disc stars, respectively, whereas 412 and 463 stars are
probabilistically thick disc and halo.  The ranges and distributions
of the galactic space velocity components for our different stellar
population categories are given in Table 2 and Fig. 7.

\begin{table}
\center 
\caption{The space velocity component ranges for population types of RAVE 
main sequence stars.}
\begin{tabular}{cccc}
\hline
Parameters &  $U$ & $V$ & $W$\\
     &  (km s$^{-1}$) & (km s$^{-1}$) & (km s$^{-1}$)\\
\hline
$TD/D \leq 0.1$     &  (-85, 75)   & (-80, 40)  & (-55, 40)  \\
$0.1 < TD/D \leq 1$ &  (-105, 100) & (-100, 50) & (-70, 50)  \\
$TD/D > 1$          &  (-165, 155) & (-175, 60) & (-115, 90) \\
\hline
\end{tabular}  
\end{table}

\begin{figure}
\begin{center}
\includegraphics[scale=0.45, angle=0]{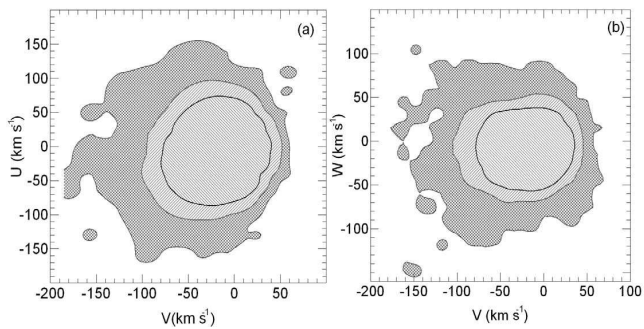}
\caption[] {Distribution of main sequence stars of different
  population types on two Galactic space velocity planes: (a) $U$-$V$ and
  (b) $W$-$V$. The contours indicate (outwards from the centre) high
  probability thin disc ($TD/D$ $\leq$ 0.1), low probability thin disc
  (0.1 $<$ $TD/D$ $\leq$ 1) and low and high probability thick disc
  and halo stars ($TD/D$ $>$ 1).}
\end{center}
\end{figure}

\section{Results}

We analysed the $U$, $V$ and $W$ space velocity components for the main
sequence sample (20453 stars) as well as for four subsamples, i.e.
high probability thin disc stars of different spectral types a) F, G
and K ($N$=17889 stars), b) F ($N$=9654 stars), c) G ($N$=5910 stars),
d) K ($N$=2325 stars) and estimated the modes of the Gaussian
distributions for each category and each space velocity
component. Spectral types were determined according to the 2MASS colours
of the stars \citep[][see Table 3]{Bilir08}. The colour range is
0.08$<$$J-H$$\leq$0.30 for F spectral type and 0.30$<$$J-H$$\leq$0.42
for G spectral type.  The modes of these kinematic distribution
functions are the best estimators of the (negative) value of the solar motion
relative to the LSR values for the stellar group in question. The
histograms for the space velocity components for the main sequence
star sample are given in Fig. 8. Gaussian fits, which are clearly an
adequate description of the data, reveal
$U_{\odot}$=8.83$\pm$0.24, $V_{\odot}$=14.19$\pm$0.34 and
$W_{\odot}$=6.57$\pm$0.21 km s$^{-1}$ as the modes of the
corresponding histograms. The modes of the histograms for the space
velocity components for high probability thin disc stars of F, G and K
spectral types, obtained by fitting them to the Gaussian distributions
(Fig. 9), are a little different than the preceding set,
i.e. $U_{\odot}$=8.50$\pm$0.29, $V_{\odot}$=13.38$\pm$0.43 and
$W_{\odot}$=6.49$\pm$0.26 km s$^{-1}$. The largest difference in
$V_{\odot}$ can be confirmed by comparing Fig. 8b and Fig. 9b (see
Section 7 below for details).

\begin{table*}
\center 
\caption{Space velocity components of the Sun with respect to the LSR
  for five population subsamples, as described in the text.}
\begin{tabular}{lccccr}
\hline
Parameters &  Colour Range & $U$ & $V$ & $W$ &  $N$\\
     &  & (km s$^{-1}$) & (km s$^{-1}$) & (km s$^{-1}$) & \\
\hline
All sample     &  0.05$\leq$$(J-H)_0$$\leq$0.55 & 8.83 $\pm$ 0.24 & 14.19 $\pm$ 0.34 & 6.57 $\pm$ 0.21 & 20 453\\
$TD/D\leq0.1$  &  0.05$\leq$$(J-H)_0$$\leq$0.55 & 8.50 $\pm$ 0.29 & 13.38 $\pm$ 0.43 & 6.49 $\pm$ 0.26 & 18 026\\
F Spectral Type&  0.08$<$$(J-H)_0$$\leq$0.30 & 8.35 $\pm$ 0.36 & 13.14 $\pm$ 0.43 & 6.24 $\pm$ 0.27 &  9 654\\ 
G Spectral Type&  0.30$<$$(J-H)_0$$\leq$0.42 & 9.25 $\pm$ 0.50 & 14.42 $\pm$ 0.57 & 6.67 $\pm$ 0.38 &  5 910\\ 
K Spectral Type&  0.42$<$$(J-H)_0$$\leq$0.55 & 7.01 $\pm$ 0.67 & 11.96 $\pm$ 0.66 & 7.03 $\pm$ 0.38 &  2 325\\
\hline
\end{tabular}  
\end{table*}

\begin{figure*}
\begin{center}
\includegraphics[scale=0.80, angle=0]{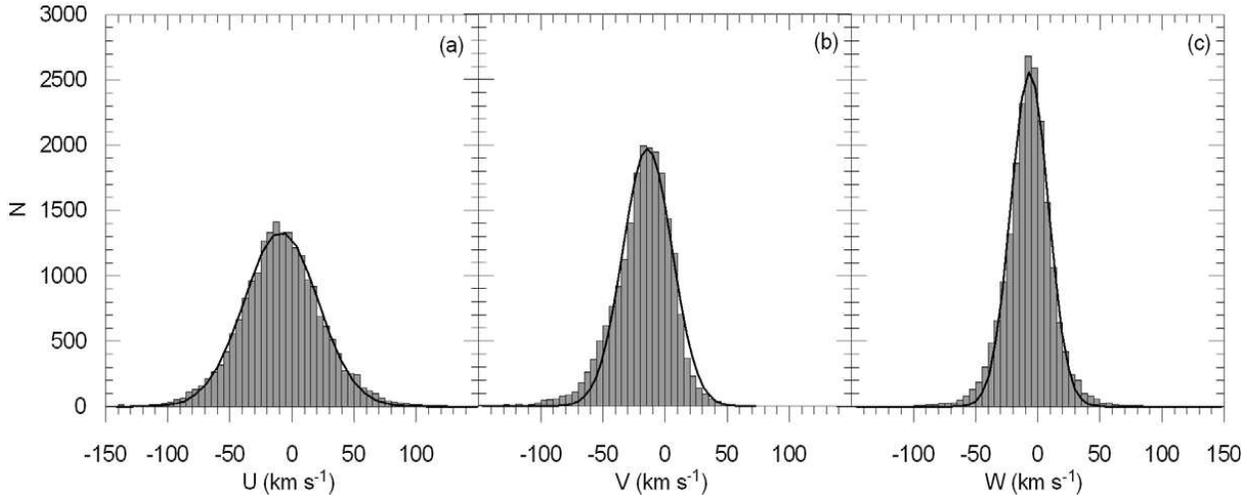}
\caption[] {The distribution functions of Galactic space velocities
  for the 20453 main sequence stars, with overlaid best-fit Gaussian distributions.}
\end{center}
\end{figure*}

\begin{figure*}
\begin{center}
\includegraphics[scale=0.80, angle=0]{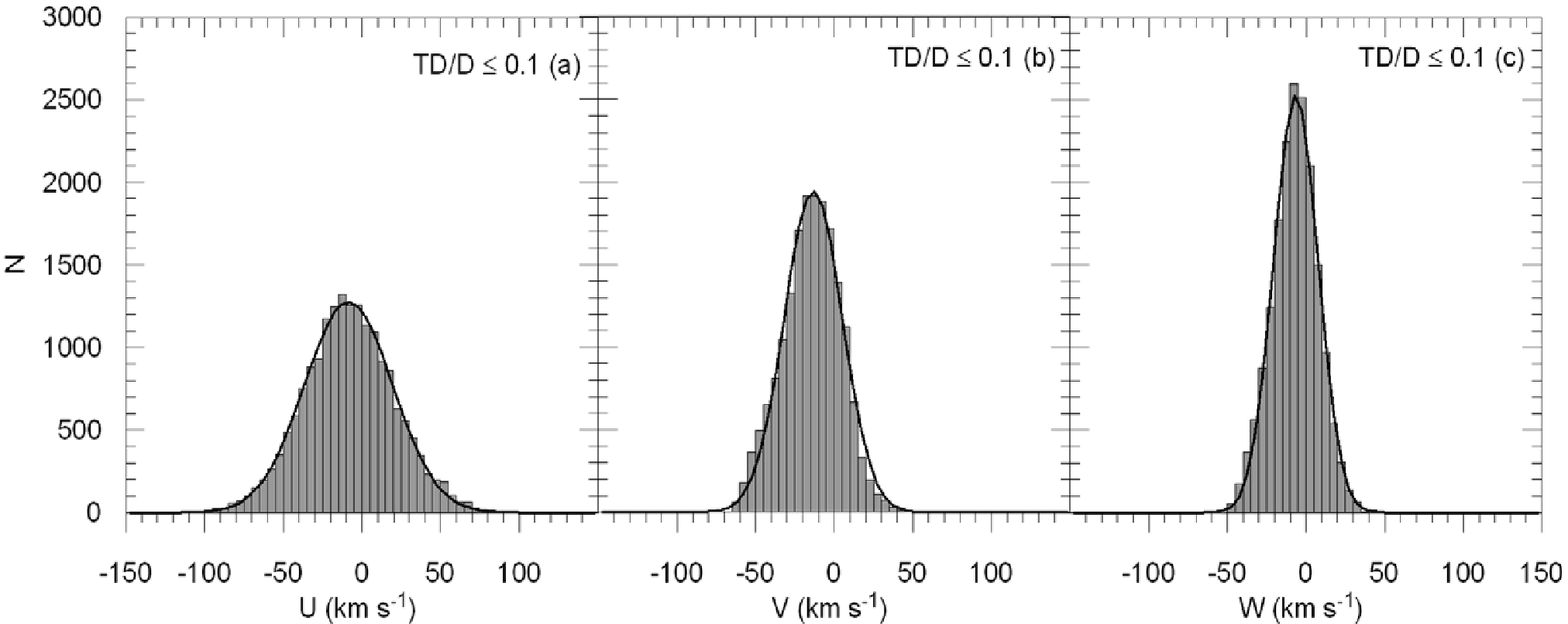}
\caption[] {The distribution functions of Galactic space velocities
  for the high probability thin disc stars, with overlaid best-fit Gaussian
  distributions.}
\end{center}
\end{figure*}

The distribution functions of Galactic space velocity components for
high-probability thin disc stars, using our kinematic population
assignment process, and presented separately for each of three
spectral-type groups, are given in Table 3. We evaluated the modes
for each distribution from the best-fit Gaussian. These results are also
given in Table 3, together with those for the high-probability main
sequence star sample and for the full sample. There are differences
between the modes of the corresponding space velocity components for
stars with F, G and K spectral types as well as between these modes
and the corresponding ones for other sub-samples which will be
discussed in Section 7.

\section{Discussion}

By definition, the Local Standard of Rest (LSR) is the reference frame
of a hypothetical star at the galacto-centric distance of the Sun, and
in the Galactic plane, that would be on a circular orbit in the
gravitational potential of the Galaxy. We can attribute $U$, $V$ and
$W$ velocity components of the Sun relative to this hypothetical star
by observing the kinematic distribution function of stars at the solar
neighbourhood, provided the stellar comparison sample we use is in
dynamical equilibrium. We can similarly test the equilibrium state of
stellar populations by comparing their deduced solar peculiar motion,
or equivalently their mean motion relative to a LSR. One may wonder if
the kinematically cold thin disc stars feel a large asymmetric
gravitational potential from, perhaps, an inner Galactic bar. Do
spiral arms have systematic kinematic effects visible in young stars?
Is the LSR defined by the high velocity halo stars, which feel largely
extended dark matter potentials, the same as that of thick disc stars,
and thin disc stars? These questions essentially repeat a query on the
closeness of the higher-order Oort Constants to zero, for all Galactic
components.  

The significance is such one wishes to determine the ``solar peculiar
velocity'' for many samples, using several techniques. Doing this (see
Table~1 above for a list) has led, in general to determinations of the
peculiar motion radially outwards from the Galactic centre $U_{\odot}$
which are consistent and small, independent of the comparison sample
chosen. Every stellar sub-population in the Galaxy seems remarkably
radially stable and circularly symmetric, with no evidence for
bar-like perturbations. Similarly, doing this for the net vertical
motion, $W_{\odot}$ provides the same conclusion. The sky fails to
fall (see also Seabroke et al., 2008). The Galaxy, locally, seems not be significantly affected by an
accreted stellar system, or asymmetric mass infall. The situation is
less clear in the rotational direction, $V_{\odot}$. Here, there is
less agreement between determinations, largely since model dependent
assumptions need to be made to correct kinematics for energy balance,
which transfers angular momentum systematically into random motions
with time. This may happen by slow diffusion, or by more dramatic
radially-dependent effects. The relevant way forward is to determine
$V_{\odot}$ as a function of as many definable ``populations'' as is
feasible, in many ways. We follow that approach here, where we find,
in agreement with several recent studies, a value of $V_{\odot}$ near
13 km s$^{-1}$. We determine this value with respect to three
sub-groups of local main-sequence stars. These groups cover a very 
wide range of stellar ages. Yet they agree, statistically, with 
each other, and with other determinations based on more local samples. 
The concept of a LSR again seems well-defined.

We study stars which are thin disc dwarfs like the Sun.
We expect a dispersion in the distribution of the velocity components
due to their inherent distributions and differences between the same
velocity components of sample stars of different population types and
ages. The way to reveal the corresponding velocity component is to fit
the space motion data to a suitable distribution function. Encouragingly, 
a simple Gaussian provides an excellent approximation so we can 
identify the mean solar motion with its mode. This is the procedure 
we used in this work.

The star sample was taken from RAVE, which extends much beyond the
distance of the high-precision local {\em Hipparcos} sample, i.e. 97\%
of the sample lies within the distance interval 0$<$$d$$\leq$0.6
kpc. We applied the following constraints to obtain a main sequence
sample with relatively small errors: i) we selected stars with surface
gravity 4$<$$\log g$$<$5, ii) we omitted stars with $(J-H)_0$$<$0.05
and $(J-H)_0$$>$0.55 to avoid blue horizontal branch and red giant
stars, iii) we excluded stars with space velocity errors larger than
25 km s$^{-1}$, iv) we separated stars probabilistically according to
their population types, and we used the high-probability thin disc
sample to minimize contamination by thick disc/halo stars as a
preferred sample in our work. We estimated solar space velocity
components for five samples: for all our RAVE main sequence stars; for
the sub-sample of high-probability thin disc main sequence stars;
and separately for the thin disc main sequence stars of F, G and K spectral
types. The results (see Table 3 above) are consistent, within sampling
errors, across all sub-samples. They are also in agreement with recent
determinations using other population subsamples, and over very
different ranges. There is increasing agreement that the solar
peculiar velocity in the rotation direction has been underestimated in
older studies. The internal agreement between our different samples,
and between this study and others, provides evidence
that the Galactic potential near the Sun is symmetric, with no evident
time dependent variation.

\section{Acknowledgments}

We thank the anonymous referee for his/her comments. Funding for RAVE 
has been provided by: the Australian Astronomical Observatory, 
the Astrophysical Institute Potsdam, the Australian
National University, the Australian Research Council, the French
National Research Agency, the German Research Foundation, the Istituto
Nazionale di Astrofisica at Padova, The Johns Hopkins University, the
W.M. Keck Foundation, the Macquarie University, the Netherlands
Research School for Astronomy, the Natural Sciences and Engineering
Research Council of Canada, the Slovenian Research Agency, the Swiss
National Science Foundation, the Science \& Technology Facilities
Council of the UK, Opticon, Strasbourg Observatory, and the
Universities of Groningen, Heidelberg, and Sydney.

This work has been supported in part by the Scientific 
and Technological Research Council (T\"UB\.ITAK) 108T613. 
Salih Karaali is grateful to the Beykent University for financial
support. This publication makes use of data products from the Two
Micron All Sky Survey, which is a joint project of the University of
Massachusetts and the Infrared Processing and Analysis
Center/California Institute of Technology, funded by the National
Aeronautics and Space Administration and the National Science
Foundation.

This research has made use of the SIMBAD, NASA's Astrophysics Data
System Bibliographic Services and the NASA/IPAC ExtraGalactic Database
(NED) which is operated by the Jet Propulsion Laboratory, California
Institute of Technology, under contract with the National Aeronautics
and Space Administration.

\end{document}